\documentstyle[prd,aps,epsfig,preprint,tighten]{revtex}
\begin{document}
\preprint{                                                BARI-TH/296-98}
\draft
\title{     Discriminating MSW solutions to the solar neutrino problem \\
            with flux-independent information at SuperKamiokande and SNO	}
\author{         G.~L.~Fogli, E.~Lisi, and D.~Montanino			}
\address{   Dipartimento di Fisica and Sezione INFN di Bari, 		\\
                  Via Amendola 173, I-70126 Bari, Italy			}
\maketitle
\begin{abstract}
%...........................................................................
The two possible Mikheyev-Smirnov-Wolfenstein (MSW) solutions of
the solar neutrino problem (one at small and the other at large mixing
angle), up to now tested mainly through absolute neutrino flux measurements, 
require flux-independent tests both for a decisive confirmation and for 
their discrimination. To this end, we perform a joint analysis of 
various flux-independent observables that can be measured at the 
SuperKamiokande and Sudbury Neutrino Observatory (SNO) experiments.
In particular, we analyze the recent data collected at 
SuperKamiokande after 374 days of operation, work out the corresponding 
predictions for SNO, and study the interplay between SuperKamiokande 
and SNO observables. It is shown how, by using only flux-independent 
observables from SuperKamiokande and SNO, one can discriminate between 
the two MSW solutions and separate them from the no oscillation case.
%...........................................................................
\end{abstract}

%%%%%%%%%%%%%%%%%%%%%%%%%%%%%%%%%%%%%%%%%%%%%%%%%%%%%%%%%%%%%%%%%%%%%%%%%%%%

	The observed difference between experimental 
\cite{Da94,Fu96,Ab96,Ha96,To97} and theoretical \cite{BP95,Ca97} absolute 
fluxes of solar neutrinos  \cite{Ba89} can be explained by the 
Mikheyev-Smirnov-Wolfenstein  (MSW) mechanism  of matter-enhanced flavor 
oscillations of neutrinos  \cite{MSWs}. Assuming for simplicity two-family 
oscillations, the MSW effect  requires a neutrino  square mass difference 
$\delta m^2\sim 10^{-5}$ eV$^2$, the mixing amplitude in vacuum 
$\sin^2 2\theta$  being either small  ($\sim 0.007$) or large ($\sim 0.7$) 
(see \cite{Li97,Fo97,Ha97,Ba97} for recent global  analyses). The nuclear 
input uncertainties affecting the calculation of the absolute $\nu$ fluxes 
\cite{Ba89,Fo95} make it desirable, however, to look also for flux-independent 
tests of the MSW effect for a definitive confirmation. Such tests can be 
performed in the new-generation, high-statistics solar $\nu$ experiments 
SuperKamiokande \cite{To97} and Sudbury Neutrino Observatory (SNO) \cite{Mc96}.

	In this work, we study the interplay among five flux-independent 
quantities measurable at SuperKamiokande and SNO, as listed in 
Table~I. The first is the asymmetry between nighttime (N) and daytime (D)
solar neutrino fluxes $(N-D/N+D)$ at SuperKamiokande, which parametrizes 
possible time variations of the $\nu$ flux due to Earth matter
effects (see \cite{Li97} and references therein). The second is the 
fractional deviation of the average kinetic energy  $\langle T \rangle$ of 
recoil electrons at SuperKamiokande ($\Delta \langle T \rangle/\langle T 
\rangle$), which parametrizes possible distortions of the neutrino energy 
spectrum due to oscillations \cite{Kr97,Li97,Fo97}. The third and fourth 
observables are the analogous quantities ($N-D/N+D$ and $\Delta \langle T 
\rangle/\langle T \rangle$) for SNO. The fifth is the ratio of charged current 
$(CC)$ to neutral current $(NC)$ neutrino event rate in deuterium $(CC/NC)$, 
which is peculiar to the SNO experiment (see \cite{Li97,Ba96}
for recent analyses). None of these quantities depends on the absolute
value of the $^8$B $\nu$ flux.

	In order to study  how the flux-independent observables can 
effectively help to test and discriminate the small (S) and large (L) mixing 
MSW solutions, we use as coordinates the observables themselves,
rather than the  usual mass-mixing parameters $\delta m^2$ and 
$\sin^2 2\theta$. In this way, the discriminative power of the 
SuperKamiokande and SNO experiments, as well as their interplay, can be 
shown and understood at glance.%
%-------------------------------------
\footnote{	An analogous choice has been made in \protect\cite{KwRo} 
		to study the relation between {\em flux-dependent\/} 
		observables at SuperKamiokande and SNO.}
%--------------------------------------
 In addition, we make use of the recent data collected
from SuperKamiokande after 374 days of operation, which have been presented 
in several recent conferences \cite{To97,Na98,In97,Su97,It97,So98,Sv97}.

	Let us start with an update of the MSW fit to the neutrino rates
measured by the Homestake \cite{La97}, Kamiokande  \cite{Fu96}, SAGE 
\cite{Ga97}, GALLEX \cite{Hp97}, and SuperKamiokande (374 days) \cite{To97}. 
The most recent available  data are compiled in Table~II. The theoretical 
ingredients and the SuperKamiokande technical specifications (energy 
resolution and threshold) are taken as in \cite{Fo97}. In particular, the 
theoretical neutrino fluxes  are taken from \cite{BP95}, the Earth regeneration 
effect is treated as in \cite{Li97}, and the experimental and theoretical 
uncertainties are included as in \cite{Fo95}. The results are shown in
Fig.~1(a) in the plane of the oscillation parameters (for simplicity, 
we consider two neutrino families). The two familiar MSW solutions, one at
small (S) and the other at large (L) mixing angle, are shown as
as allowed regions at 95\% C.L.\ ($\Delta \chi^2 = 5.99$ for 
$N_{\rm DF}=2$).
Notice that in Fig.~1(a) the fit includes 
{\em only\/} the total neutrino rates, i.e., only the
flux-dependent information. The shapes of the S and L solutions
can still change, to some extent, as this information is continually updated. 
However, they have proven to be quite stable. Therefore, 
it makes sense  to use the flux-dependent solutions S and L as a guide 
to evaluate  
the likely expectations for the flux-independent observables, as we do 
in the following.

	Next we consider the two SuperKamiokande observables
$N-D/N+D$ and $\Delta \langle T \rangle/\langle T \rangle$. The night-day
asymmetry measured in 374 days is \cite{To97}:
%..............................................................................
\begin{eqnarray}
\frac{N-D}{N+D}\times 100	&=&	3.1 \pm 2.4~{\rm (stat.)}~\pm1.4~
					{\rm (syst.)}\nonumber \\
               			&=&	3.1 \pm 2.8~{\rm (1\sigma\ total)}\ ,
\label{ND}
\end{eqnarray}
%.............................................................................
compatible with zero at $\sim 1\sigma$.  We do not use here the additional 
SuperKamiokande information  on nighttime rates in ``fractions of nadir 
angle'' \cite{In97}, since they still have a low statistical significance 
and their systematics are not available at present.

	The second SuperKamiokande flux-independent
observable, namely, the  deviation $\Delta \langle T \rangle/\langle T \rangle$ 
of the average (measured) electron kinetic energy  from its standard value, 
can be obtained from the  binned energy spectrum (after 374 days 
\cite{To97,Sv97,Priv})  reported in Table~III, by following the
prescription described in the Appendix of Ref.~\cite{Fo97}. 
The result is:
%.......................................................................
\begin{eqnarray}
\frac{\Delta\langle T\rangle}{\langle T\rangle}\times 100 
 	=     0.95 &\pm& 0.62~{\rm (stat.+uncor.~syst.)}\nonumber\\
     	&\pm& 0.30~{\rm (correl.~syst.)}\nonumber\\
     	&\pm& 0.20~({}^8{\rm B~theor.~shape)}\nonumber\\
     	&\pm& 0.12~{\rm (binning)}\nonumber\\
 	=     0.95 &\pm& 0.73~{\rm (1\sigma\ total)}\ .
\label{dT/T}
\end{eqnarray}
%........................................................................
In the above equation, the first error is due to statistics and to those 
systematics which are not correlated bin-by-bin (e.g., backgrounds). The 
second error is due to the those systematics that are fully correlated in 
each bin, the most important being the energy scale uncertainty 
\cite{In97,Ba96}. The third error is due to the theoretical uncertainties in 
the shape of the $^8$B neutrino spectrum \cite{Bspe}. Finally, the fourth 
uncertainty is due to the finite bin size \cite{Fo97} (it could be avoided if
the value of ${\Delta\langle T\rangle}/{\langle T\rangle}$ were
estimated directly from the raw data by the SuperKamiokande collaboration).
From Eq.~(2) we learn that: 
	(i) 	there seems to be a slight $(\sim 1.3\,\sigma)$
	  	enhancement of the average electron energy, that might 
		signal a distortion of  the energy spectrum; 
	(ii) 	at present, the largest error component is due
		to statistics plus uncorrelated systematics; and 
	(iii) 	the $^8$B $\nu$ spectrum uncertainty is nonnegligible.

	In Fig.~1(b) the small (S) and large (L) mixing solutions of Fig.~1(a)
are mapped in the plane spanned by  the flux-independent SuperKamiokande 
observables $N-D/N+D$ and $\Delta \langle T \rangle/\langle T \rangle$. The 
small box with error bars represents the experimental
data reported in Eqs.~(1) and (2). Also shown for comparison are the older 
SuperKamiokande data (black dot with error bars) after 306 days of operation 
(see \cite{Fo97} and refs.\ therein), indicating a dramatic reduction of the 
electron energy uncertainty after 374 days, 
as a result of an improved
energy calibration of the detector \cite{In97}. The two MSW solutions are 
well separated in the plane of Fig.~1(b). In fact, the small mixing solution 
predicts $\Delta \langle T \rangle/\langle T \rangle \simeq 0.8$--$1.4\%$ and
$N-D/N+D \simeq 0$--$4\%$, while the large mixing solution 
corresponds to smaller values of $\Delta \langle T \rangle/\langle T \rangle$ (
in the range $\pm 0.3\%$) and prefers relatively large values of
$N-D/N+D$ (in the range  $\simeq 2$--$17\%$).

	From Fig.~1(b) it can be seen that the flux-independent
SuperKamiokande data  favor the small mixing solution, although the
experimental uncertainties are not small enough yet to exclude either the large mixing
solution or the no oscillation point [star in Fig.~1(b)].
Figure~1(b) also shows that, in order to discriminate the two MSW solutions 
between them and from the no oscillation scenario, the SuperKamiokande error 
bars should be reduced by a factor of two at least (which seems a
reachable, although difficult, goal \cite{In97}).  
In particular, improvements in the experimental 
determination of $\Delta \langle T \rangle/\langle T \rangle$ appear decisive
for separating the S and L solutions (assuming no significant changes in the 
experimental central values).

	Next we study the relation between the SuperKamiokande and SNO
flux-independent observables, which is the main goal of this work. 
For SNO we take the same prospective technical specifications as in 
\cite{Ba96,Li97}; in particular, we assume an electron kinetic energy 
threshold $T>5$ MeV and 100\%  detection efficiency. 
With these assumptions, one has  $(CC/NC)_0=1.88$ for no oscillation 
\cite{Ba96,Li97}; this number should be rescaled appropriately when the SNO 
efficiency will be measured.

	Figure~2 shows SuperKamiokande vs.\ SNO flux-independent
observables in various combinations. In each panel we map the two
MSW solutions (S and L), as well as the no oscillation point (star).
The solutions are well separated in the panels (a--d), while in the panels 
(e) and (f)  there is some overlap. There is a high correlation between the 
expected values of $N-D/N+D$ in SuperKamiokande and SNO [Fig.~2(e)], which
are approximately in a 1:2 ratio. Similarly, the $\Delta \langle T \rangle/
\langle T \rangle$ values in Fig.~2(a)
are highly correlated, with about the same proportionality 
factor. This means that, given a sufficiently precise measurement at 
SuperKamiokande {\em and\/} assuming the correctness of the MSW hypothesis, 
one can derive rather definite predictions for SNO.

	As an application, we study the implication of the data in Eqs.~(1) 
and (2), which, in Fig.~2, select the horizontal dotted bands at the
the $\pm 1\sigma$ level. The corresponding predictions for SNO can be 
roughly obtained by projecting the part of the MSW solutions {\em contained\/} 
within the  SuperKamiokande bands onto the $x$-axes, and taking their 
intersection. For instance, from the panels (a) and (d)
one derives $\Delta\langle T\rangle/\langle T\rangle\simeq 1.8$--$4.5$ 
as the likely $\pm1\sigma$ range for SNO (notice that the $x$-axis
projections of the region L in panels (a) and (d) do not intersect).
We can do an analogous exercise for panels 2(b,e) and 2(c,f). In summary,
one obtains the following ``$\sim 1\sigma$ predictions'' for SNO:
%..........................................................................
\begin{equation}
{\rm SKam}\left\{ 
\begin{array}{ccl}
\frac{N-D}{N+D}&=&3.1\pm2.8\% \\
\frac{\Delta\langle T\rangle}{\langle T\rangle}&=&0.95\pm 0.73\% \\
\end{array}\right.\ +{\rm MSW}
\ \Longrightarrow\ 
{\rm SNO}\;(\sim1\sigma)\, \left\{ 
\begin{array}{ccl}
\frac{N-D}{N+D}&\simeq&0{\rm-}6\% \\
\frac{\Delta\langle T\rangle}{\langle T\rangle}&\simeq&1.8{\rm -}4.5\% \\
\frac{CC/NC}{(CC/NC)_0}&\simeq&15{\rm -}50\%\ .
\end{array}\right.
\end{equation}
%..........................................................................
Of course, if the SuperKamiokande data in the above equation
were used at $>1\sigma$ level, then the predictive power for SNO would be 
rapidly lost, as a consequence of the 
relatively large experimental uncertainties.

	If real SNO data were available, one could  also draw in Fig.~2
``allowed vertical bands'' whose intersection with the horizontal bands
should then spot one of the two MSW solutions.  The panels in Fig.~2 allow one 
to set easily the experimental accuracy needed to separate the 
MSW solutions between them and from the no oscillation point,
using exclusively flux-independent information from SNO and SuperKamiokande. 
For instance, if the SuperKamiokande errors on 
${\Delta\langle T\rangle}/{\langle T\rangle}$ and $N-D/N+D$ were reduced by a 
factor of two at least, and if the corresponding SNO measurements could reach 
a similar fractional accuracy, then the S and L solutions would
be easily distinguished, although with a statistical significance depending 
on the specific position of the data points. In particular, in
the unlucky case of data points incidentally falling close to
the no oscillation point (star), Figs.~2(a,b,d,e) show that the
discriminative power would be reduced. In such cases (and in
more general situations as well) the $CC/NC$ measurement at SNO 
provides invaluable help, as evident from Figs.~2(c,f). In fact,
with a prospective $CC/NC$ uncertainty of a few percent, the
two MSW solutions would be clearly separated from the no oscillation
case (see also Ref.~\cite{Ba96} for earlier studies). Therefore,
using various combinations of flux-independent observables from
SuperKamiokande and SNO, one should be able to separate the MSW solutions 
between them and from the no oscillation case. The experimental accuracy 
needed for such discrimination  (which depends, to some extent, on the 
central value of the data) can be easily evaluated through Fig.~1(b) 
and Fig.~2 by drawing prospective error bands for SNO and SuperKamiokande.

	In conclusion, we have shown the interplay between SuperKamiokande
and SNO flux-independent observables within the MSW interpretation of the
solar neutrino deficit. We have updated the small and large mixing angle
MSW solutions by including the most recent flux-dependent data, and
used such solutions as a guide for the flux-independent analysis.
The SuperKamiokande energy spectrum and day-night asymmetry  data have also 
been used to derive predictions for SNO. The estimated likely ranges for SNO
are still rather large, but will steadily narrow as the SuperKamiokande
uncertainties get reduced. Moreover, when the SNO experiment will also
start operation, various combinations of SuperKamiokande and SNO
flux-independent data will allow to spot either the large or the
small mixing MSW solution, and to separate them from the no oscillation
case. Our graphical representations show at glance the interplay between
the two experiments, and allow to estimate {\em a priori\/} the
experimental accuracy needed to separate the various cases with an
assigned statistical significance.

%%%%%%%%%%%%%%%%%%%%%%%%%%%%%%%%%%%%%%%%%%%%%%%%%%%%%%%%%%%%%%%%%%%%%%%%%%%%

One of us (E.L.) thanks the organizers of the  Solar Neutrino Workshop 
``News about SNU's'' (Institute of Theoretical Physics, Santa Barbara, CA),
where this work was initiated, for their kind hospitality  and support.

%%%%%%%%%%%%%%%%%%%%%%%%%%%%%%%%%%%%%%%%%%%%%%%%%%%%%%%%%%%%%%%%%%%%%%%%%%
%		               T A B L E S
%%%%%%%%%%%%%%%%%%%%%%%%%%%%%%%%%%%%%%%%%%%%%%%%%%%%%%%%%%%%%%%%%%%%%%%%%%

%%%%%%%%%%%%%%%%%%%%%%%%%%%%%%%%%%%%%%%%%%%%%%%%%%%%%%%%%%%%%%%%%%%%%%%%%%
%%%%%%%%%%%%%%%%%%%%%%%%  TABLE I  %%%%%%%%%%%%%%%%%%%%%%%%%%%%%%%%%%%%%%%
\begin{table}
\caption{	The five flux-independent observables
		considered in this work.}
\begin{tabular}{lll}
%=========================================================================
 Experiment 	& Observable	& Definition 	\\
\tableline
%-------------------------------------------------------------------------
 SuperKamiokande	& $(N-D)/(N+D)$	& Night-day flux asymmetry \\
 		& $\Delta \langle T\rangle/\langle T\rangle $ 
 &Energy spectrum deviation\\
\tableline
%-------------------------------------------------------------------------
SNO		& $(N-D)/(N+D)$	& Night-day flux asymmetry \\
		& $\Delta \langle T\rangle/\langle T\rangle $ 
 &Energy spectrum deviation\\
		& $CC/NC$	& Charged-to-neutral current ratio		
%=========================================================================
\end{tabular}
\end{table}

%%%%%%%%%%%%%%%%%%%%%%%%%%%%%%%%%%%%%%%%%%%%%%%%%%%%%%%%%%%%%%%%%%%%%%%%%%
%%%%%%%%%%%%%%%%%%%%%%%%  TABLE II  %%%%%%%%%%%%%%%%%%%%%%%%%%%%%%%%%%%%%%%
\begin{table}
\caption{	Neutrino event rates measured by solar neutrino experiments,
		and corresponding predictions from the  standard solar model
		\protect\cite{BP95}. The quoted errors are at $1\sigma$.}
\begin{tabular}{ccccc}
%=========================================================================
Experiment 	&Ref.& Data~$\pm$(stat.)~$\pm$(syst.)
& Theory \protect\cite{BP95}& Units \\
\tableline
%-------------------------------------------------------------------------
Homestake	& \protect\cite{La97}& $ 2.56 \pm 0.16 \pm 0.15$         &
	$9.3^{+1.2}_{-1.4}$    & SNU					\\
Kamiokande 	& \protect\cite{Fu96}& $ 2.80 \pm 0.19 \pm 0.33$         &
	$6.62^{+0.93}_{-1.12}$ & $10^6$ cm$^{-2}$s$^{-1}$		\\
SAGE		& \protect\cite{Ga97}&$69.9^{+8.5}_{-7.7}{}^{+3.9}_{-4.1}$&
	$137^{+8}_{-7}$        & SNU					\\
GALLEX		& \protect\cite{Hp97}& $76.4 \pm  6.3^{+4.5}_{-4.9}$     &
	$137^{+8}_{-7}$	       & SNU					\\
SuperKam.	& \protect\cite{To97}&$2.37^{+0.06}_{-0.05}{}^{+0.09}_{-0.06}$&
 	$6.62^{+0.93}_{-1.12}$ & $10^6$ cm$^{-2}$s$^{-1}$	
%=========================================================================
\end{tabular}
\end{table}

%%%%%%%%%%%%%%%%%%%%%%%%%%%%%%%%%%%%%%%%%%%%%%%%%%%%%%%%%%%%%%%%%%%%%%%%%%%
%%%%%%%%%%%%%%%%%%%%%%%  TABLE III  %%%%%%%%%%%%%%%%%%%%%%%%%%%%%%%%%%%%%%%%
\begin{table}
\caption{Energy spectrum of recoil electrons measured at SuperKamiokande
	(374 d). The numbers in columns 2--4 have been graphically reduced 
	from the binned spectrum in \protect\cite{To97,Priv} and represent:
	The ratio between data and theory (2nd column); the corresponding
	$\pm 1\sigma$ errors due to statistics plus uncorrelated systematics
	(3rd column) and to correlated systematics (4th column).
	 Small asymmetries between upper and lower error bars have been 
	 neglected.} 
\begin{tabular}{cccc}
%=========================================================================			&
Bin energy 		&
expt./theor.		&
stat.+unc.~syst.	&
correl.~syst.		\\
range (MeV) 		&
spectrum ratio		&
uncertainties		&
uncertainties		\\
\tableline
%-------------------------------------------------------------------------
$[6.5,\,7]$ 	& 0.327	& $\pm0.030$ 	& $\pm0.004$	\\
$[7,\,7.5]$ 	& 0.335 & $\pm0.029$ 	& $\pm0.006$	\\
$[7.5,\,8]$ 	& 0.402 & $\pm0.030$ 	& $\pm0.008$	\\
$[8,\,8.5]$ 	& 0.354 & $\pm0.032$ 	& $\pm0.010$	\\
$[8.5,\,9]$ 	& 0.304 & $\pm0.032$ 	& $\pm0.012$	\\
$[9,\,9.5]$ 	& 0.345 & $\pm0.036$ 	& $\pm0.014$	\\
$[9.5,\,10]$	& 0.382 & $\pm0.042$ 	& $\pm0.016$	\\
$[10,\,10.5]$	& 0.367 & $\pm0.045$ 	& $\pm0.017$	\\
$[10.5,\,11]$	& 0.308 & $\pm0.045$ 	& $\pm0.023$	\\
$[11,\,11.5]$	& 0.377 & $\pm0.055$ 	& $\pm0.026$	\\
$[11.5,\,12]$	& 0.360 & $\pm0.063$ 	& $\pm0.036$	\\
$[12,\,12.5]$	& 0.404 & $\pm0.077$ 	& $\pm0.042$	\\
$[12.5,\,13]$	& 0.390 & $\pm0.087$ 	& $\pm0.046$	\\
$[13,\,13.5]$	& 0.504 & $\pm0.119$ 	& $\pm0.063$	\\
$[13.5,\,14]$	& 0.490 & $\pm0.146$ 	& $\pm0.101$	\\
$[14,\,20]$ 	& 0.562 & $\pm0.154$ 	& $\pm0.114$
%=========================================================================
\end{tabular}
\end{table}

%%%%%%%%%%%%%%%%%%%%%%%%%%%%%%%%%%%%%%%%%%%%%%%%%%%%%%%%%%%%%%%%%%%%%%%%%%%
% 			R E F E R E N C E S 
%%%%%%%%%%%%%%%%%%%%%%%%%%%%%%%%%%%%%%%%%%%%%%%%%%%%%%%%%%%%%%%%%%%%%%%%%%%

%\end{document}

%%%%%%%%%%%%%%%%%%%%%%%%%%%%%%%%%%%%%%%%%%%%%%%%%%%%%%%%%%%%%%%%%%%%%%%%%%%
%		F I G U R E S 
%%%%%%%%%%%%%%%%%%%%%%%%%%%%%%%%%%%%%%%%%%%%%%%%%%%%%%%%%%%%%%%%%%%%%%%%%%%

%..........................................................................
\begin{figure}
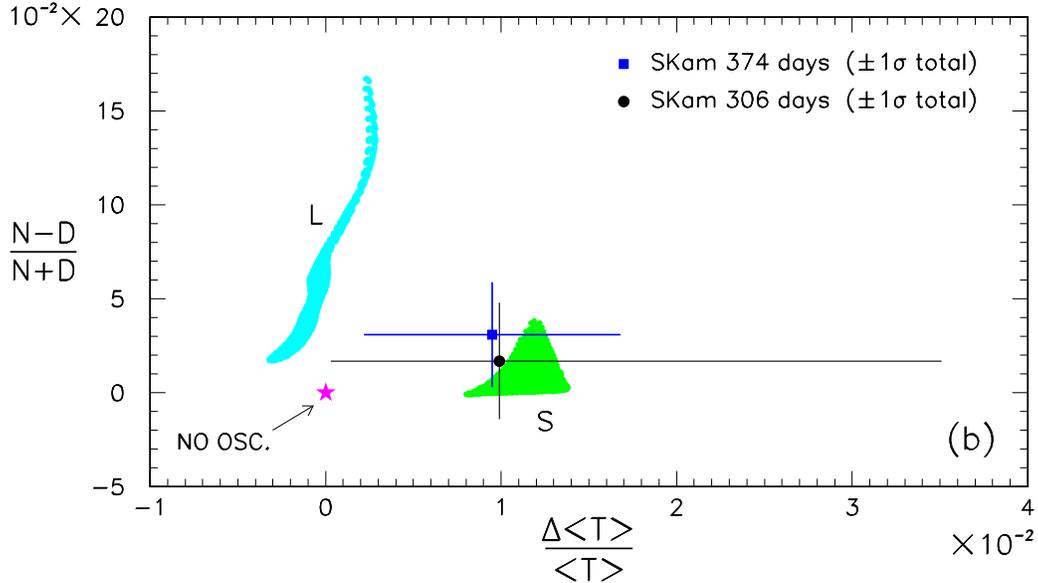

\caption{	Summary of flux dependent and independent information
		on the MSW solutions. (a) MSW fit to the
		total neutrino rates measured by SuperKamiokande (374 days),
		Kamiokande, Homestake, SAGE and GALLEX (see Table~I),
		showing the small mixing (S) and large mixing (L) solutions
		at 95\% C.L. in the neutrino mass-mixing plane. (b) Map 
		of the S and L solutions in the plane spanned by two
		flux-independent SuperKamiokande observables: the day-night
		asymmetry $N-D/N+D$ and the mean kinetic energy deviation
		$\Delta \langle T \rangle/\langle T \rangle$. Also shown
		are the experimental data after 306 and 374 days 
		of operation, and the ``no oscillation'' point. See the
		text for details.}
\label{fig:1}
\end{figure}
%..........................................................................
\begin{figure}
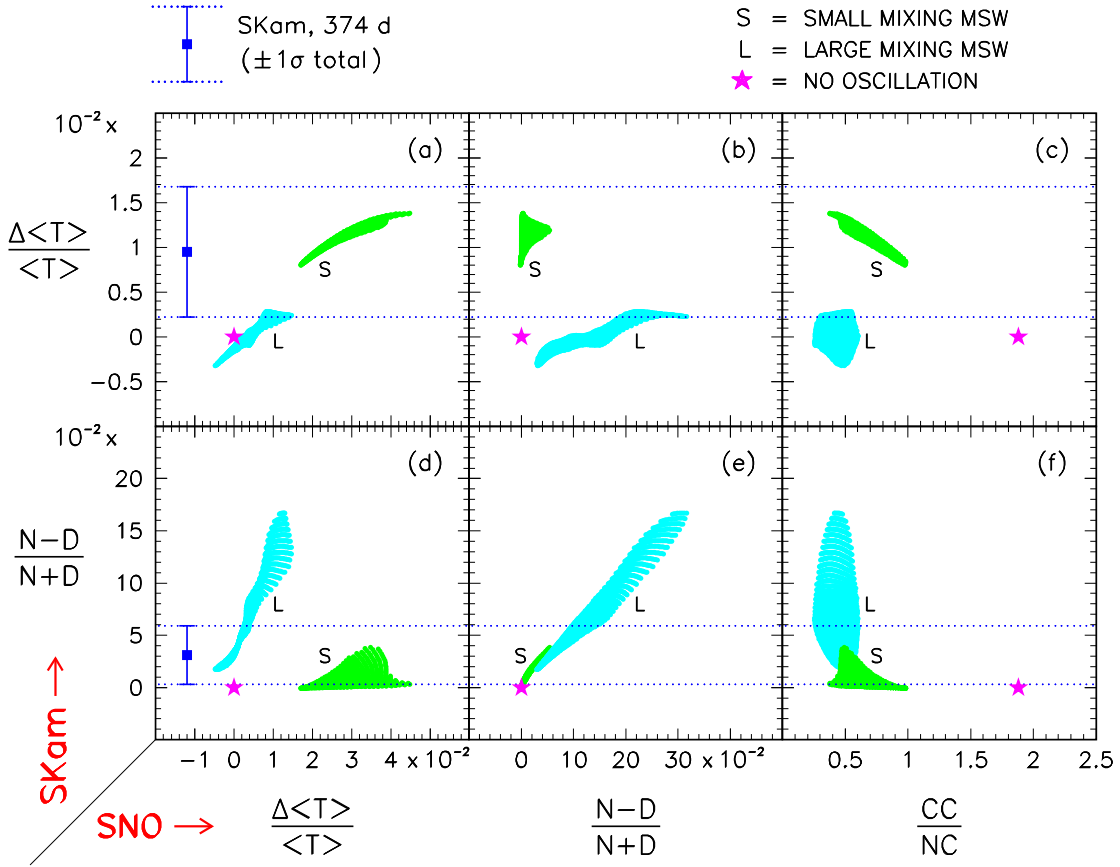

\caption{	Relation between flux-independent observables at
		SuperKamiokande ($\scriptstyle \frac{N-D}{N+D}$ and
		$\scriptstyle\frac{\Delta \langle T \rangle}{\langle T \rangle}$
		and at SNO ($\scriptstyle\frac{N-D}{N+D}$,
		$\scriptstyle\frac{\Delta \langle T \rangle}{\langle T 
		\rangle}$, and $\scriptstyle\frac{CC}{NC}$). 
		The S and L solutions of Fig.~1(a) are mapped
		onto each plane, together with the "no oscillation" point
		(star). The horizontal bands represent the $\pm 1\sigma$
		data from SuperKamiokande (374 days). 
		The MSW predictions for SNO (at $\sim 1\sigma$ level)
		can be obtained by projecting the part of the
		MSW solutions contained within such bands onto
		the $x$-axes, and taking their intersection.}
\label{fig:2}
\end{figure}
%..........................................................................

%\end{document}

%%%%%%%%%%%%%%%%%%%%%%%%%%%%%%%%%%%%%%%%%%%%%%%%%%%%%%%%%%%%%%%%%%%%%%%%%%%
%%%%%%%%%%%%%%%%%%%%%%%%%%%%%%%%%%%%%%%%%%%%%%%%%%%%%%%%%%%%%%%%%%%%%%%%%%%
%%%%%%% 
%%%%%%%            THE FOLLOWING FOR AUTHOR USE ONLY.
%%%%%%%
%%%%%%%            INCLUSION OF FIGURES WITH EPSFIG.STY.
%%%%%%%
%%%%%%%%%%%%%%%%%%%%%%%%%%%%%%%%%%%%%%%%%%%%%%%%%%%%%%%%%%%%%%%%%%%%%%%%%%
%%%%%%%          P O S T S C R I P T       F I G U R E S 
%%%%%%%
%%%%%%%   memo:  1) add epsfig in the \documentstyle
%%%%%%%          2) and move this part before \end{document} 
%%%%%%%		 3) remove previous figure captions
%%%%%%%          4) include the following \newcommand:
%%-------------------------------------------------------------------------
\newcommand{\InsertFigure}[2]{\newpage\begin{center}\mbox{%
\epsfig{bbllx=1.4truecm,bblly=1.3truecm,bburx=19.5truecm,bbury=26.5truecm,%
height=21.truecm,figure=#1}}\end{center}\vspace*{-2.3truecm}%
\parbox[t]{\hsize}{\small\baselineskip=0.5truecm\hskip0.5truecm #2}}
%--------------------------------------------------------------------------
%%%%%%%%%%%%%%%%%%%%%%%%%%%%%%%%%%%%%%%%%%%%%%%%%%%%%%%%%%%%%%%%%%%%%%%%%%%

%..........................................................................
\InsertFigure{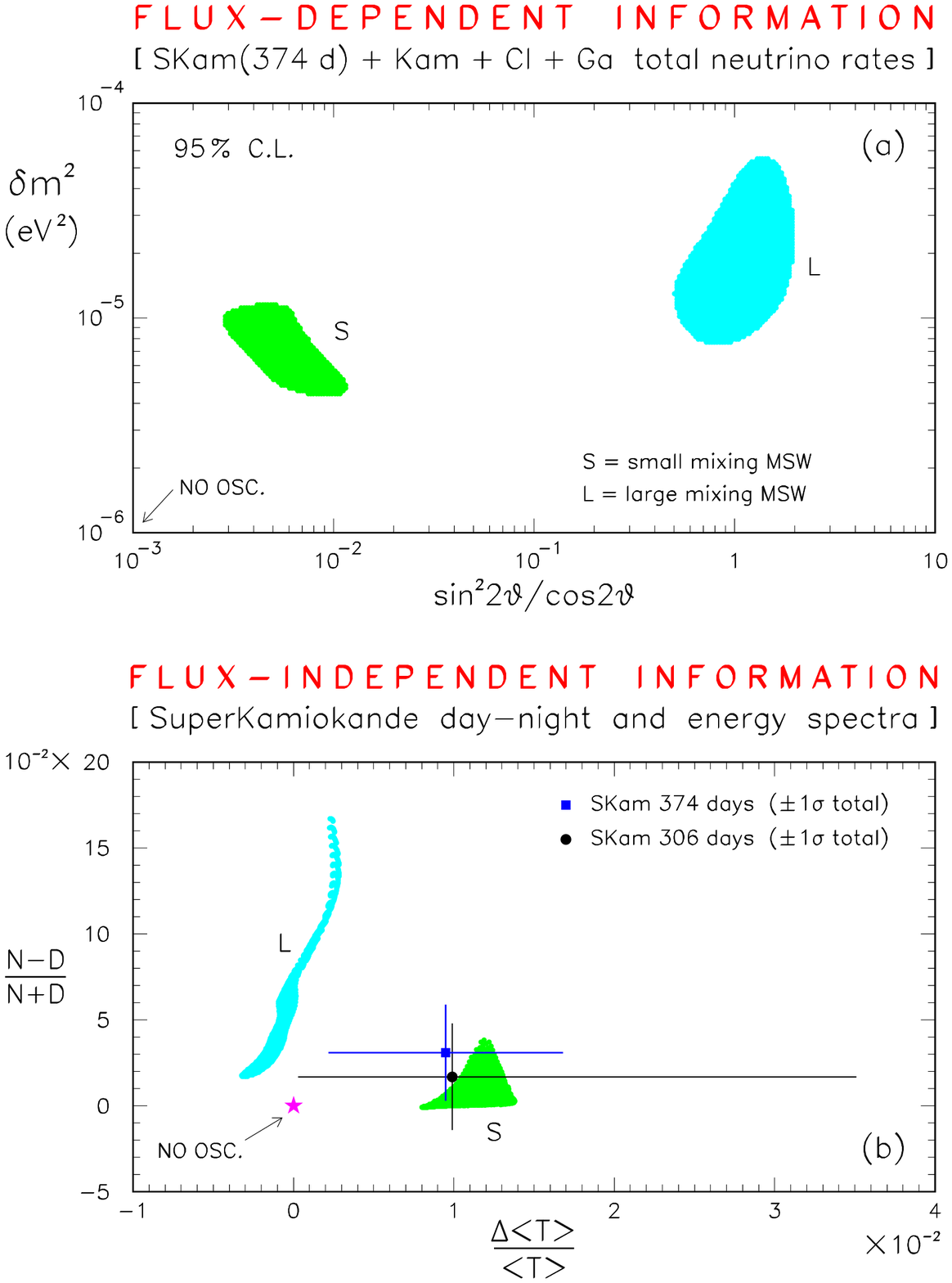}%
{FIG.~1.	Summary of flux dependent and independent information
		on the MSW solutions. (a) MSW fit to the
		total neutrino rates measured by SuperKamiokande (374 days),
		Kamiokande, Homestake, SAGE and GALLEX (see Table~I),
		showing the small mixing (S) and large mixing (L) solutions
		at 95\% C.L. in the neutrino mass-mixing plane. (b) Map 
		of the S and L solutions in the plane spanned by two
		flux-independent SuperKamiokande observables: the day-night
		asymmetry $N-D/N+D$ and the mean kinetic energy deviation
		$\Delta \langle T \rangle/\langle T \rangle$. Also shown
		are the experimental data after 306 and 374 days 
		of operation, and the ``no oscillation'' point. See the
		text for details.}
%..........................................................................
\InsertFigure{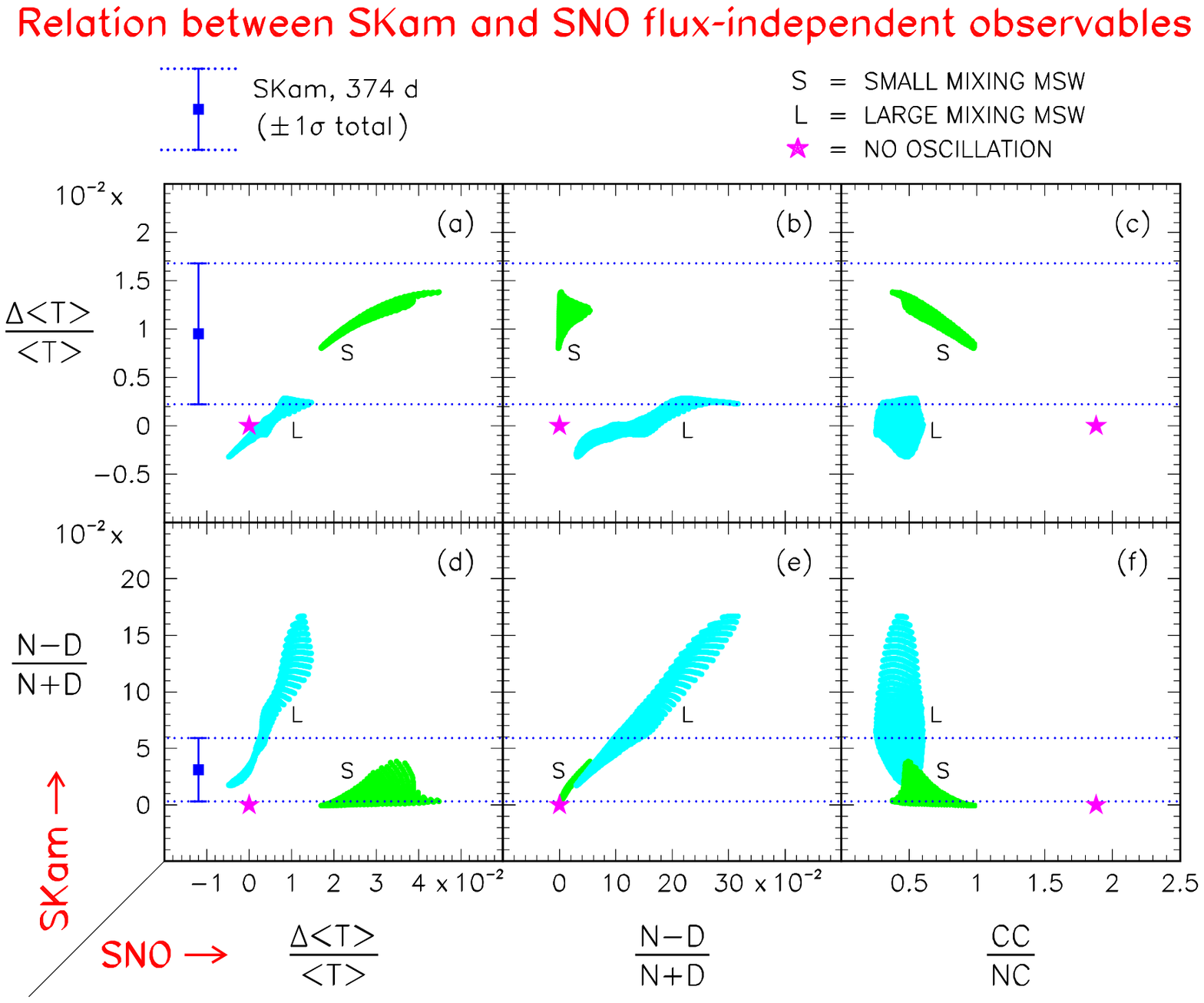}%
{FIG.~2. 	Relation between flux-independent observables at
		SuperKamiokande ($\scriptstyle \frac{N-D}{N+D}$ and
		$\scriptstyle\frac{\Delta \langle T \rangle}{\langle T \rangle}$
		and at SNO ($\scriptstyle\frac{N-D}{N+D}$,
		$\scriptstyle\frac{\Delta \langle T \rangle}{\langle T 
		\rangle}$, and $\scriptstyle\frac{CC}{NC}$). 
		The S and L solutions of Fig.~1(a) are mapped
		onto each plane, together with the "no oscillation" point
		(star). The horizontal bands represent the $\pm 1\sigma$
		data from SuperKamiokande (374 days). 
		The MSW predictions for SNO (at $\sim 1\sigma$ level)
		can be obtained by projecting the part of the
		MSW solutions contained within such bands onto
		the $x$-axes, and taking their intersection.}	
%..........................................................................

\end{document}